\documentclass[conference]{IEEEtran}
\usepackage{graphicx}
\usepackage{amsmath}
\usepackage{amssymb}
\usepackage{color}
\usepackage{ifpdf}
\usepackage{float}
\usepackage[utf8]{inputenc}
\usepackage{multirow}
\usepackage{rotating}
\usepackage{subfigure}

\usepackage{moresize}
\usepackage{url}
\usepackage{booktabs}
\usepackage{listings}   
\usepackage{paralist}    
\usepackage{wrapfig}    
\usepackage{multirow}
\usepackage{ifpdf}
\usepackage{xspace}
\usepackage{keyval}  
\usepackage{color}

\definecolor{listinggray}{gray}{0.95}
\definecolor{darkgray}{gray}{0.7}
\definecolor{commentgreen}{rgb}{0, 0.4, 0}
\definecolor{darkblue}{rgb}{0, 0, 0.4}
\definecolor{middleblue}{rgb}{0, 0, 0.7}
\definecolor{darkred}{rgb}{0.4, 0, 0}
\definecolor{brown}{rgb}{0.5, 0.5, 0}
\definecolor{dkgreen}{rgb}{0,0.5,0}
\definecolor{orange}{rgb}{1,.5,0}
\definecolor{dandelion}{cmyk}{0,0.29,0.84,0}

\usepackage[normalem]{ulem}
\makeatletter
\def\cyanuwave{\bgroup \markoverwith{\lower3.5\p@\hbox{\sixly \textcolor{cyan}{\char58}}}\ULon}
\def\reduwave{\bgroup \markoverwith{\lower3.5\p@\hbox{\sixly \textcolor{red}{\char58}}}\ULon}
\def\blueuwave{\bgroup \markoverwith{\lower3.5\p@\hbox{\sixly \textcolor{blue}{\char58}}}\ULon}
\font\sixly=lasy6 
\makeatother

\newif\ifdraft
\ifdraft
 \newcommand{\N}[1]{\textbf{NOTE: #1}\xspace}
 \newcommand{\jhanote}[1]{ {\textcolor{red} { ***SJ: #1 }}}
 \newcommand{\katznote}[1]{ {\textcolor{blue} { ***DSK: #1 }}}
 \newcommand{\mtnote}[1]{ {\textcolor{orange} { ***MT: #1 }}}
 \newcommand{\jonnote}[1]{ {\textcolor{dkgreen} { ***JW: #1 }}}
 \newcommand{\mingnote}[1]{ {\textcolor{magenta} { ***MTH: #1 }}}
 \newcommand{\note}[1]{ {\textcolor{brown} { *** #1 }}}
 \newcommand{\vbnote}[1]{ {\textcolor{ForestGreen} { ***VB: #1 }}} 
\else
 \newcommand{\N}[1]{}
 \newcommand{\jhanote}[1]{}
 \newcommand{\katznote}[1]{}
 \newcommand{\mtnote}[1]{}
 \newcommand{\jonnote}[1]{}
 \newcommand{\mingnote}[1]{}
 \newcommand{\note}[1]{}
 \newcommand{\vbnote}[1]{}
\fi

\newcommand{\rct}{RADICAL-Cybertools\xspace}
\newcommand{\entk}{EnTK\xspace}

\lstdefinestyle{myListing}{
  frame=single,   
  backgroundcolor=\color{listinggray},  
  language=C,       
  basicstyle=\ttfamily \footnotesize,
  breakautoindent=true,
  breaklines=true
  tabsize=2,
  captionpos=b,  
  aboveskip=0em,
  belowskip=-2em,
}      

\lstdefinestyle{myPythonListing}{
  frame=single,   
  backgroundcolor=\color{listinggray},  
  language=Python,       
  basicstyle=\ttfamily \scriptsize,
  breakautoindent=true,
  breaklines=true
  tabsize=2,
  captionpos=b,  
}

\ifpdf
\DeclareGraphicsExtensions{.pdf, .jpg, .tif}
\else
\DeclareGraphicsExtensions{.ps,  .eps, .jpg}
\fi

\begin{document}

\title{Designing Workflow Systems Using Building Blocks}

\author{Matteo Turilli$^{1}$, Andre Merzky$^{1}$, Vivek
Balasubramanian$^{1}$, Manuel Maldonado$^{1}$, Shantenu Jha$^{1}$$^{,2}$\\
   \small{\emph{$^{1}$ Rutgers, the State University of New Jersey, Piscataway, NJ 08854, USA}}\\
   \small{\emph{$^{2}$ Brookhaven National Laboratory, Upton, New York,
11973}}\\ }

\date{}
\maketitle

\begin{abstract} 
We suggest there is a need for a fresh perspective on the design and
development of workflow systems and argue for a building blocks approach. We
outline a description of this approach and define the properties of software
building blocks. We discuss RADICAL-Cybertools as one implementation of the
building blocks concept, showing how they have been designed and developed in
accordance with this approach. Four case studies are presented, covering a
dozen science problems. We discuss how RADICAL-Cybertools have been used to
develop new workflow systems capabilities and integrated to enhance existing
ones, illustrating the applicability and potential of software building
blocks. In doing so, we have begun an investigation of an alternative
approach to thinking about the design and implementation of workflow systems.
\end{abstract}

\section{Introduction}\label{sec:intro}

Sophisticated and scalable workflows have come to epitomize advances in
computational science, especially for ``big science'' projects, such as those
in high-energy physics or astronomy. Workflows are also becoming more
pervasive across application types, scales and communities.

Interestingly, many commonly used workflow systems in high-performance and
distributed computing such as Kepler~\cite{Ludascher2006},
Pegasus~\cite{Deelman2005}, and Swift~\cite{zhao2007swift} emerged from an
era when the software landscape supporting distributed computing was fragile,
missing features and services. Not surprisingly, these initial and many
subsequent workflow systems, had a monolithic design that included the
end-to-end capabilities needed to execute workflows on heterogeneous and
distributed cyberinfrastructures.

In spite of the many successes of workflow systems, there is a perceived
barrier-to-entry and limited flexibility. There continues to be an absence of
a reasoning framework for end-users to determine which systems to use, when
and why. For example, such a lack of clarity and guidance was cited as the
single most pressing barrier to workflow adoption by participants of a recent
Blue Waters Workflows Workshop. These issues in part are a consequence of the
lack of a structured approach to system design and of an unsustainable
fragmented ecosystem comprised of systems that often need to establish
exclusivity (i.e., ``vendor lock-in'') or by preserving ``domains of
influence''.

Without negating monolithic workflow systems where the socio-technical needs
warrant them and make their use meaningful, a valid question is whether it is
possible to construct workflow systems or extend the available ones avoiding
some of the aforementioned shortcomings. This question is set against the
increasing richness of workflow-based applications and consequent demands on
workflow systems. Is it possible to design these systems to provide greater
flexibility and sharing of features while not constraining functionality,
performance, or sustainability?

An important but often overlooked fact is that the majority of scientific
workflows don't use existing workflow systems. The reasons are varied and are
not just limited to the proverbial ``last mile customization'' challenge of
workflows. A corollary to the above observation is that there is a need for a
sustainable ecosystem of software components from which tailored workflow
systems can be composed, as opposed to having to fit workflows to
pre-existing solutions. Thus, the challenge is to support the development and
composition of workflow systems that can be responsive to the wide range of
workflow requirements. This challenge supersedes the need to develop a
software to substitute all other workflow systems, or to interoperate with
all of them.

Several additional factors motivate a discussion of alternative approaches to
the design and engineering of workflow systems. These systems need to be
better prepared for new application scenarios (e.g., integration of
large-scale experiments, instruments and observation devices with
high-performance computing), scale (e.g., exascale high-performance
computing), while improving our ability to create lower-cost and sustainable
solutions. Furthermore, the variety and importance of applications with a
large number of possibly concurrent simulations are growing while, at the
same time, the software platforms and services available in support of
science has improved in both robustness and features.

This paper makes the case for taking a fresh perspective to the design,
development and integration of workflow systems by means of a building blocks
approach. In the next section we describe this approach and its four design
principles of self-sufficiency, interoperability, composability, and
extensibility.  We postulate the building blocks approach overcomes the
limited flexibility of monolithic workflow systems without the significant
burden typically associated with integrating disparate software systems.  We
also argue that the building block approach is a better fit for the typical
academic development and economic model, and that developing software
building blocks complements the existing systems by helping to avoid software
duplication, resource starvation and lack of functionalities.

Section~\ref{sec:rct} discusses how we used the building blocks approach to
design and develop \rct. These are a set of software modules that can be used
independently, composed into a single system, or integrated into existing
systems to extend their functionalities. We introduce a four-layered view of
high-performance and distributed systems and we describe how each software
module implements distinctive functionalities for each layer.

Section~\ref{sec:casestudies} discusses four case studies of employing \rct
as building blocks to develop or integrate workflow systems. The first study
describes the creation of domain-specific workflow systems tailored to the
biomolecular and seismology domains. The other three studies illustrate how
the functionalities of mainstream workflow systems can be extended by
integrating \rct. The case studies cover different types of applications,
ranging from distributed high-throughput analysis jobs on single nodes
(PanDA), to multiple simulations (SeisFlows, HTBAC, FireWorks) to adaptive
workflows of biomolecular simulations (RepEx, ExTASY).

We conclude with a discussion of the practical impact of the case studies as
well as the lessons learnt by testing the validity and feasibility of the
building blocks approach. We highlight the benefits of implementing new
capabilities into existing workflow systems by integrating the \rct. We also
outline the limitations of our contributions as well as some open questions.

\textbf{Terminology:} The term `workflow' is often overloaded in literature
and used for a wide variety of scenarios in the computational science
discourse. Sometimes, the term workflow is used to describe the computational
process associated with an application; sometimes to indicate the application
itself. To add to the confusion, `workflows' are sometimes also used as a
reference to the task graph (i.e., tasks and their relationships)
representing the application. Even when there is clarity on what a workflow
describes, a complicating factor is that there are multiple distinct
specifications of the same workflow.

Another common source of confusion is the conflation between similar but
distinct concepts, such as those of workflow and workload. In this paper, we
adopt the following definitions: A multi-task application can be represented
as a {\bf workflow}, i.e., a set of tasks with dependencies that determine
the order of their execution. Subsets of these tasks can be {\bf workloads},
i.e., tasks whose dependencies have been satisfied at a particular point in
time and that may be executed concurrently. In this way, workflow provides a
complete description of the execution process while workload identifies the
entity that is executed. We maintain that these characteristics are
independent of the scale of the application, the number of users (or
developers) of the workflow or type of workflow (compute or data-intensive).
As such, Workflow systems and Workload systems control different entities.

Although the focus of this paper is on using the building blocks approach to
design, develop, and integrate workflow systems, the approach we propose is
equally applicable to workload management systems, prominent examples of
which are PanDA~\cite{maeno2008panda},
glideinWMS~\cite{sfiligoi2008glideinwms} or DIRAC~\cite{stagni2012lhcb}.

\section{Related Work}\label{sec:related}

We classify existing workflow systems into three categories, focusing only on
those with the highest adoption and ongoing development. All-inclusive
workflow systems such as Kepler~\cite{Ludascher2006},
Swift~\cite{wilde2011swift}, Fireworks~\cite{jain2015fireworks} and
Pegasus~\cite{Deelman2005} provide full-featured end-to-end capabilities that
includes application creation, execution, monitoring and provenance. General-
purpose workflow systems such as Ruffes~\cite{goodstadt2010ruffus},
COSMOS~\cite{gafni2014cosmos}, and GXP make~\cite{taura2013design} also
enable end-to-end execution but prioritize the simplicity of their
interfaces, limiting the range of capabilities. Finally, domain-specific
workflow systems such as Galaxy~\cite{goecks2010galaxy},
Taverna~\cite{oinn2004taverna}, BioPipe~\cite{hoon2003biopipe}, and
Copernicus~\cite{pronk2015molecular} focus on providing interfaces tailored
to the requirements of specific domain scientists.

Decomposition of workflow systems into self-contained components fosters the
decoupling of software development efforts that can be performed
independently using standardized interfaces. Self-contained components are
simpler to understand and enables participation of developers with diverse
skills and leads to a bigger developer community. This improves both
maintainability and extensibility. Systems can be composed by plugging in
different implementations of these components based on the user requirements
and the component features as the interface is standardized and
implementation details are abstracted. SAGA~\cite{saga-x} is a software tool
that enables interoperability over multiple heterogeneous computing
infrastructures via an API standardized by the Open Grid Forum. Multiple
scientific applications and workflow
systems~\cite{iwai2010saga,frank2007interoperable,lordan2014servicess,lezzi2012interoperable,huang2010magate}
that require resource interoperability have been developed with SAGA as the
common tool.

Similarly, we motivate the decomposition of workflow systems into components
with high intra-component cohesion and low inter-component dependency. These
components themselves can be implemented in monolithic or modular fashion,
but with well-defined objectives and standardized interfaces multiple
workflow systems can be built depending on the requirements of the users.

Modularity, in software deployment, has evolved from chroot~\cite{chroot},
jails~\cite{kamp2000jails}, and solaris zones~\cite{tucker2004solaris} into
modern day micro-services~\cite{dragoni2017microservices} and other service
oriented architectures. These approaches evolve from the concepts of
component-based software engineering~\cite{heineman2001component,
crnkovic2011software} (CBSE). It is important to note that, in this paper, we
do not suggest reinvention of existing CBSE concepts, but point to the
benefits of its application to workflow systems for scientific computing.

\section{Building Block Approach}\label{sec:bba}

The building block approach is related to the methods presented in
Ref.~\cite{batory1992design,lenz1988software,garlan1995architectural}. In
this paper, we apply this approach to the design of middleware  for the
execution of scientific workflows. In our adaptation, the building block
approach is used to describe the architectural design  of workflow systems
and is based on four design principles: self-sufficiency, interoperability,
composability, and extensibility.

A software building block is self-sufficient when its design does not depend
on the specificity of other building blocks; interoperable when it can be
used in diverse system architectures without semantic modifications;
composable when its interfaces enable communication and coordination with
other building blocks; and extensible when the building's block
functionalities and entities can be extended to support new requirements or
capabilities.

For example, a system component designed to handle only a single type of
workflow does not satisfy the principle of self-sufficiency. Analogously, a
system capable of managing multiple types of workflows but only when
expressed in a specific representation does not satisfy the principle of
interoperability. Software systems designed for unidirectional communication
or without the capability to enable coordination cannot be composed so as to
form a distributed system with end-to-end capabilities. Finally, systems that
cannot be extended cannot guarantee sustained interoperability and
composability.

Each building block has a set of entities and a set of functionalities that
operate on these entities. Architecturally, a building block designed in
accordance with the method we propose has: (i) two well-defined and stable
interfaces, one for input and one for output; (ii) one or more conversion
layers capable of translating across diverse representations of the same type
of entity; (iii) one or more modules implementing the functionalities to
operate on these entities. While this architecture is relatively common, it
has been seldom applied to the design of middleware systems for the execution
of scientific workflows.

Self-sufficiency and interoperability depend upon the choice of both entities
and functionalities. Entities have to be general enough so that specific
instances of that type of entity can be reduced to a unique abstract
representation. Accordingly, the scope of the functionalities of each
building block has to be limited exclusively to its entities. In this way,
interfaces can be designed to receive and send diverse codifications of the
same type of entity, while functionalities can be codified to translate
consistently those representation into a generic set of properties, and
operate on them.

Composibility depends on whether the interfaces of each building block
enables communication and coordination. Blocks communicate information about
the states, events and properties of their entities enabling the coordination
of their functionalities. Due to the requirement of self-sufficiency, the
building block approach uses an explicit model of the entities' states and
transitions. The coordination among blocks cannot be assumed to happen
implicitly ``by design''; it has to be explicitly codified on the base of a
documented state model. The sets of entities and functionalities need to be
extensible to enable the coordination among states of multiple and diverse
blocks. Note that extensibility remains bound by both interoperability and
self-sufficiency.

As seen in Section~\ref{sec:intro}, a large variety of workflow systems have
been designed as self-contained, end-to-end systems implementing specific
codifications of the workflows, dedicated engines for their execution, and
customized interfaces to specific sets and type of resources. While this
approach has often worked, especially in terms of performance and initial
``time-to-market'', it fosters duplication and thus has an impact on the
sustainability of their implementations and the functionalities available.

The building block approach avoids these traps while proposing a viable
alternative to the redesign and implementations of prototyped end-to-end
systems. The development of scientific software by academic labs is mostly
incompatible with large groups of specialized engineers and with the
socioeconomic models required for their
sustainability~\cite{baxter2006scientific,segal2008developing}. In the
presence of scarce and transient economic and development resources, but also
of stringent requirements of complexity, the building block approach
leverages the large amount of domain knowledge available in academic groups
and the benefit of well-defined but contained design and development
endeavors. It thus avoids the faux equivalence between an academic and
commercial development environment.

It is worth mentioning that we are not advocating the rejection of existing
monolithic workflow systems in favor of a building block approach. We believe
an ecosystem that facilitates the co-existence and cooperation of end-to-end
workflow systems along with well-defined building blocks is an optimal
situation. Such a heterogeneous ecosystem would be indication of good health
for the workflows community, contributing to address the challenges of the
next generation of computation and data-enabled sciences.

In the following section, we present a set of tools that have been designed
in accordance to the proposed building block approach. These tools have been
iteratively design and implemented across ten years to support not only
diverse user communities and case studies but also a principled investigation
of their own design, requirements, performance, and maintainability. Of
significance, is the broad range of production grade workflow types,
communities and scales they support. We discuss these in
Section~\ref{sec:casestudies}.

\section{RADICAL Cybertools}\label{sec:rct}

\begin{figure}[t]
  \includegraphics[width=\columnwidth]{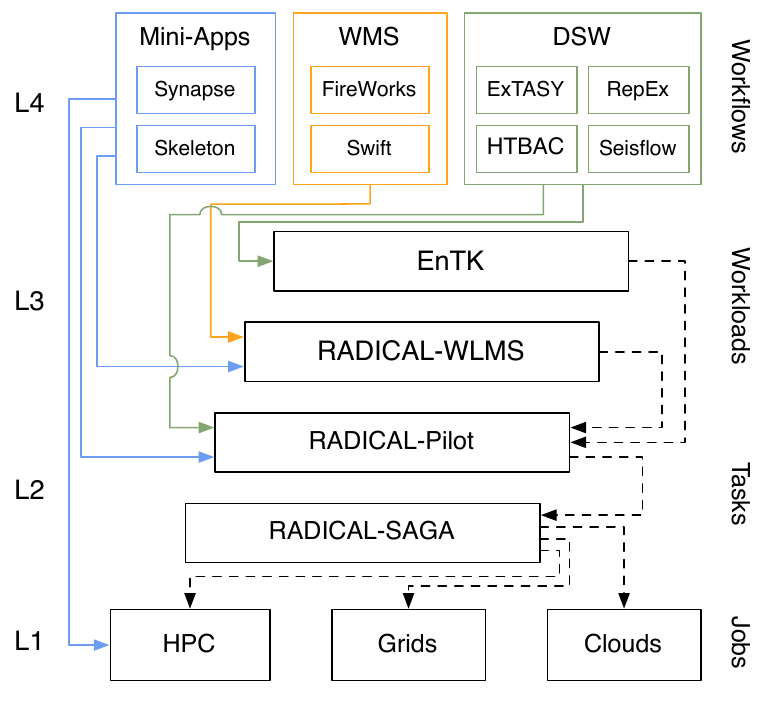}
  \caption{End to end composition of RADICAL-Cybertools. Numbered levels on
           the left; names of entities on the right. Solid colored lines
           indicate composition between workflows/applications and \rct;
           dashed lines composition among \rct. Blue, orange, and green lines
           indicate how tools,
           mini-apps~\cite{merzky2015synapse,katz2016application}, workflow
           systems and domain-specific workflows (DSW) are executed via
           alternative compositions of
           \rct.\label{fig:radical-bb}}
\end{figure}

\rct are software modules designed and implemented in accordance with the
building block approach described in Section~\ref{sec:bba}.  Each module has
been designed independently from each other and with well-defined
functionalities and entities. Fig.~\ref{fig:radical-bb} shows four \rct
modules alongside their inter-relationships: RADICAL-SAGA~\cite{saga-x},
RADICAL-Pilot~\cite{review_radicalpilot}, RADICAL-WLMS~\cite{ipdps-2016}, and
RADICAL Ensemble-Toolkit hereafter simply referred as \entk~\cite{entk}.

We briefly discuss a four-layered view of high-performance and distributed
systems as depicted in Fig.~\ref{fig:radical-bb} to appreciate the design of
individual \rct as well as their overall organization.

\begin{figure}[!ht]
  \begin{center}
  \includegraphics[width=\columnwidth]{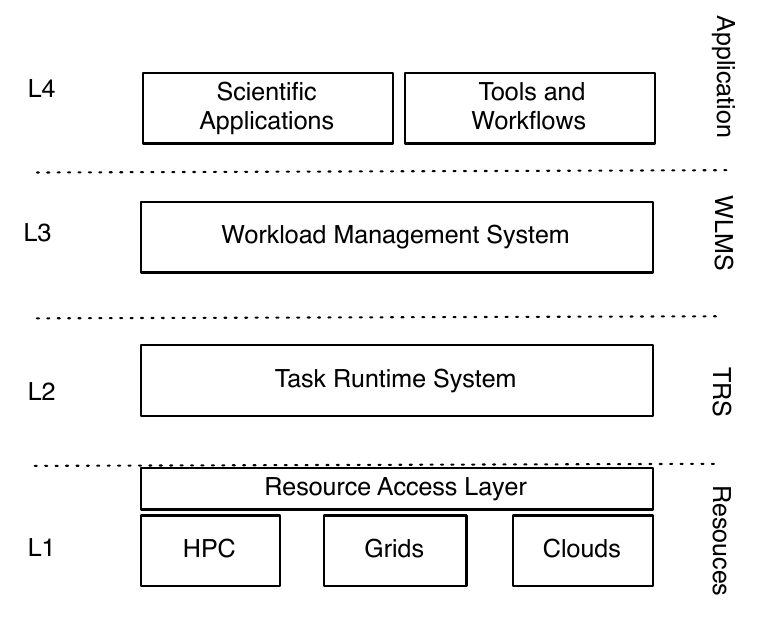}
  \caption{Primary functional levels. The diagram supports an analysis of the
           functional requirements for workflow systems, and the primary
           entities at each level, agnostic of the applications and
           resources.\label{fig:layers}}
\end{center}
\end{figure}

Each layer has a well-defined functionality and an associated ``entity''. The
entities start from \textbf{workflows} (or applications) at the top layer and
resource specific \textbf{jobs} at the bottom layer, with intervening
transitional entities of \textbf{workloads} and \textbf{tasks}. The diagram
of Fig.~\ref{fig:layers} provides a reference example for transitions between
these entities across layers that is independent of the specifics of workload
and resources.

\textit{Workflow and Application Description Level (L4)}: Provides an
expressive yet flexible way to capture the requirements and semantics of the
applications and workflows.

\textit{Workload Management (WLM) Level (L3):} Applications devoid of
semantic context are expressed as workloads, which are a set of tasks whose
relationships and dependencies are expressed as a computational graph. The
Workload Management layer is responsible for: (i) the selection and
configuration of available resources for the given workload; (ii)
partitioning the workload over the selection of suitable resource; (iii)
binding of constituent tasks to resources; and (iv) the management and
coordination of these three functional aspects.

\textit{Task Execution Runtime Level (L2):} L3 delivers tasks to L2 which is
responsible for their effective and efficient execution on the selected
resources. L2 is a passive recipient of tasks from L3 but includes an active
module that maps the tasks onto a scheduling overlay comprised of the chosen
computing resources.

\textit{Resource Layer (L1):} The resources used to execute tasks are
characterized by their capabilities, availability and interfaces. At L1, all
tasks have been wrapped up as resource specific jobs; while the semantic
inconsistency in the capabilities of resources remains, each job can be
submitted to diverse resources thanks to advances in syntactically uniform
resource access layers.

We now discuss the four \rct and how they conform to the principles of
self-sufficiency, interoperability, composability and extensibility.

RADICAL-SAGA exposes a homogeneous programming interface to the queuing
systems of HPC, HTC, and cloud resources. SAGA---an OGF
standard~\cite{saga-x}---abstracts away the specificity of each queue system,
offering a consistent representation of jobs and of the capabilities required
to submit them to the resources. The design of RADICAL-SAGA is based on the
job entity and the scope of its functionalities is limited to job submission
and jobs' requirements handling (self-sufficiency). Both entities and
functionalities can be extended to support, for example, new queue systems or
new type of jobs (extensibility). The SAGA API resolves the differences of
each queue system into a general and sufficient representation
(interoperability), exposing a stable set of capabilities to both users
and/or other software elements (composability).

RADICAL-Pilot is a pilot system implemented in accordance with the pilot
model described in Ref.~\cite{pstar12,review_pilotreview}. RADICAL-Pilot
exposes an API to enable the acquisition of resource placeholders on which to
schedule workloads for execution. This API is implemented both as a library
and as a RESTful service. The design of RADICAL-Pilot includes pilot, and
compute and data unit as entities. Capabilities are made available to
describe, schedule, manage and execute entities. Pilots, units and their
functionalities abstract the specificities of diverse type of resources,
enabling the use of pilots on single and multiple HPC, HTC, and cloud
resources. A pilot can span single or multiple compute nodes, resource pools,
or virtual machines. Units of various size and duration can be executed,
supporting MPI and non-MPI executables, with a wide range of execution
environment requirements.

The design of RADICAL-Pilot~\cite{review_radicalpilot,cug-2016} is:
self-sufficient due to the generality and well-defined scope of its entities
and functionalities; interoperable in terms of type of workload, resource,
and execution requirements; and extensible as new properties can be added to
the pilot and unit description, and more capabilities can be implemented
without altering its design.  Currently, composability is partially designed
and implemented: while the PILOT-API can be used by both users and other
systems to describe one or more generic workloads for execution,
RADICAL-Pilot interfaces to resources requires SAGA\@. A system based on
dedicated resource connectors, including but not limited to SAGA, is
currently being designed.

The design of RADICAL-WLMS is also an ongoing project. Developed as a
prototype to study and test a general model of workload management,
RADICAL-WLMS is being developed to be agnostic towards the modalities used by
users or systems to provide workloads descriptions. RADICAL-WLMS integrates
information about the workload requirements and the resource capabilities,
explicitly separating the planning and management of each workload execution.

RADICAL-WLMS uses an abstraction called ``execution strategy'' for the
homogeneous specification of alternative execution plans, and an execution
manager to enact each plan. Workloads are executed over one or more pilots,
with number of cores and duration tailored to the requirements of the
workload. Pilots are concurrently scheduled on one or more resources, and
units are scheduled concurrently into every available pilot. This enables
dynamic slicing of the workload so to optimize the size, duration, and
binding of pilots, and the placement of units on those pilots.

We have also used the building block approach to coordinate the distributed
execution of applications with specific computational patterns. \entk
promotes ensembles as a first-class entity and has the following design
features to meet the requirements of ensemble-based applications: (i) enable
the expression of an ensemble of tasks abstracting the specificity of the
tasks' executable; (ii) support for commonly used and pre-determined
ensemble-based execution patterns; (iii) decoupling of the expression of
patterns from the management of their execution; and (iv) a runtime system
that enables the efficient execution of tasks and provides flexible resource
utilization capabilities over a range of HPC platforms.

\entk adheres to the four elements of the building block approach: It is
self-sufficient as it is not limited to a specific type of ensemble or
execution pattern, and thus is fully general in the scope of its entities and
functionality. \entk is interoperable across different executables and
resources, supporting composability and extensibility by exposing an API
tailored to the development of execution patterns, some of which are
predefined for the user but which can be arbitrarily extended.

It is important to note how each RADICAL-Cybertool has been designed to be
used independently. Each cybertool is not designed as part of an overall
system: each module is a system in itself. Several independent communities
directly utilize RADICAL-SAGA without using RADICAL-Pilot, and other
communities have been using RADICAL-SAGA with alternative pilot systems
implementations. This is the essence of the building block approach we are
proposing. Progressively, each cybertool will be further developed to be used
by a diverse research communities and with diverse workflow, pilot, manager,
or broker systems.

\rct do not implement new types of system. Workflow, workload and execution
managers are common modules of many middleware supporting the distributed
execution of workloads and workflows. The novelty rests with their design
approach, not with their functionalities.  Their adoption by a wide range of
end users but also by projects that have already developed their own software
modules is a testament to the relevance of the approach here proposed. We
discuss details of their uptake in the next section.

\section{Case Studies}\label{sec:casestudies}

In this section, we discuss four case studies in which \rct have been
independently integrated with user-facing libraries and production-grade
systems developed by distinct teams at different institutions and from a
range of disciplines. Where the same \rct module have been used across
different case studies, the points  of integration  have been different.
These case studies illustrate how the building blocks approach enables
integration by implementing minimal new functionalities.

The first case study involves domain-specific workflow (DSW) systems from
biomolecular sciences and seismology integrated with the \entk module of
\rct. The second and third case studies review the integration of the two
workflow systems Swift~\cite{zhao2007swift} and
FireWorks~\cite{jain2015fireworks} with RADICAL-WLMS~\cite{ipdps-2016}. The
fourth case study describes the integration of
PanDA~\cite{angius2017converging} with the Next Generation Executor (NGE)
module.

The level at which systems are integrated differs in each case study. The
first case study integrates four DSW systems with a single workflow manager;
the second a workflow manager with an execution manager; the third an
execution manager with a resource manager; and the fourth a broker with a
pilot system. Further, each integration uses a different type of interface:
API, file system, methods, and database. Finally, each integration implements
a different element of a coordination protocol by passing tasks to a workflow
manager, a master process, a resource, or a pilot.

The entities `task', `pilot', and `resource' remain invariant across the
integrations. This avoids reimplementation of functionalities in favor of
translation layers among, for example, data structures representing tasks
properties and relations, or resource requirements and capabilities. It
should be noted that while these entities are specific to the domain of
workflow and workloads, the building block approach can be used in every
domain with a well-defined set of entities.

\subsection{DSW Systems and \entk}\label{ssec:dsw}

\begin{figure}
    \centering
    \includegraphics[width=0.49\textwidth]{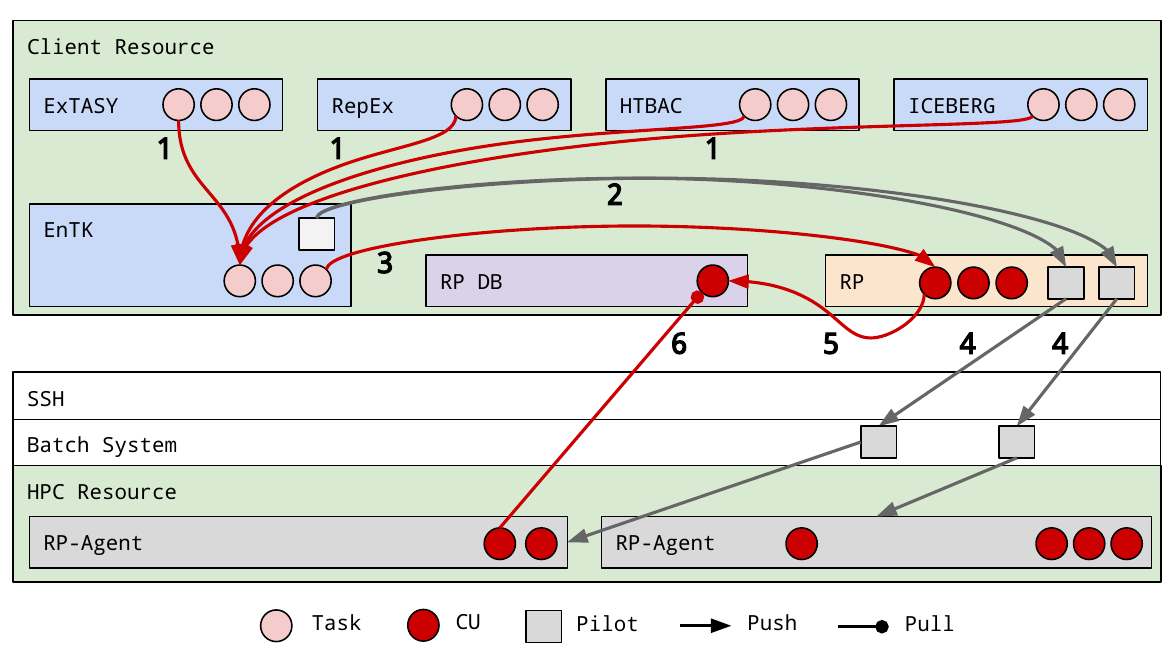}
    \caption{Integration between four domain-specific workflow systems
            (ExTASY, RepEx, HTBAC, SeisFlows) and \entk. Numbers indicate the
            execution flow. RADICAL-Pilot (RP) database (DB) can be deployed
            on any host reachable from the resources. RP does not vary across
            integrations (Fig.~\ref{fig:swift_integration},
            \ref{fig:fireworks_integration},
            \ref{fig:panda_integration}).\label{fig:entk_integration}}
\end{figure}

\entk, described in the previous section, has been used to build four DSW
systems to support workflows that are characterized by different ensemble-
based execution patterns (Figure~\ref{fig:entk_integration}). \entk is
agnostic to the details of both the specific executables run by the ensemble
and the system used to manage their execution. In
Figure~\ref{fig:entk_integration}, \entk is coupled with RADICAL-Pilot to
execute the ensembles via pilots on HTC but, in principle, \entk could use a
different runtime system.

\entk was used to build the ExTASY toolkit~\cite{extasy}, a DSW that supports
several sampling methods in biomolecular simulations. ExTASY invokes \entk as
a Python library and uses the Ensemble API to provide the \textit{simulation-
analysis} execution pattern. Several sampling algorithms (LSDMap and CoCO)
are consistent with this execution pattern and are implemented using ExTASY.

\entk also supports the \textit{replica-exchange} pattern, and is thus usable
by RepEx~\cite{repex} which is a DSW enabling multiple replica-exchange
methods. RepEx has been shown to be a powerful framework to support
multi-dimensional and exchange schemes~\cite{ct500776j}. RepEx achieves this
by separating the performance layer from the functional layer, while
providing simple and easy methods to extend interfaces.

Two additional DSW use \entk: The first is the High-throughput binding
affinity calculator (HTBAC) which is used to  determine clinically relevant
binding affinities~\cite{bac-2008}. The other is
SeisFlows~\cite{seisflows-rtd}, an open source seismic inversion package that
delivers customizable waveform inversion workflows so as to support research
in regional, global, and exploration seismology.

HTBAC implements the ESMACS and TIES protocols to calculate binding free
energies~\cite{esmacs-ties-jacs17}. These workflows consist of consecutive MD
runs (for example equilibration and production) followed by post processing
steps. Although ESMACS and TIES methods are similar at a high-level, i.e.,
they are comprised of concurrent, multi-stage pipelines with synchronization,
they differ in the details of the pipelines, stages and synchronization.
HTBAC uses the \entk API to express these workflows; \entk provides advanced
resource management capabilities and, thereby delivers the necessary
high-throughput capabilities required. \entk provides a common API, execution
and programming model to these different methods, and thus will minimize
development effort and complexity.

SeisFlows is designed to support seismic inversion workflow, at scale on HPC
machines. The workflow is comprised of multiple sequential and concurrent
stages of the workflow and associated data movement which are supported using
\entk. The associated tool---SeisFlows---is used for fast prototyping of
seismic workflows uses RADICAL-SAGA to extract information from a database to
execute jobs. It is mostly used to run data pre-processing and simulations
sub-workflows.

All four DSW systems (ExTASY, RepEx, HTBAC and SeisFlows) benefit from the
use of \entk by not having to reimplement workload management, efficient task
management and interoperable task execution on distinct and heterogeneous
platforms. This in turn enables both a focus on and ease of ``last mile
customization'' for the DSW\@.

\subsection{Swift and RADICAL-WLMS}\label{ssec:swift}

For our study, we choose Swift, which is both a language and a runtime system
to specify and execute workflows. Swift has a long development history, with
several versions that supported diverse case studies. Swift also integrated
pilot systems of which Coasters~\cite{hategan2011coasters} is actively
supported. The design of Swift is modular and it relies on connectors to
interface with third-party systems.

In Swift, the language interpreter and the workflow engine are tightly
coupled but connectors can be developed to stream the tasks of workflows to
other systems for their execution. As seen in the previous section, all \rct
can get streams of tasks as an input: RADICAL-SAGA can submit these tasks as
jobs to diverse resources; RADICAL-Pilot can schedule these tasks into
pilots; and RADICAL-WLMS can derive and enact a suitable execution strategy
to execute the given tasks. Each \rct offers a different and well-isolated
set of capabilities, depending on the specific set of abstractions they
implement.

\begin{figure}
    \centering
    \includegraphics[width=0.49\textwidth]{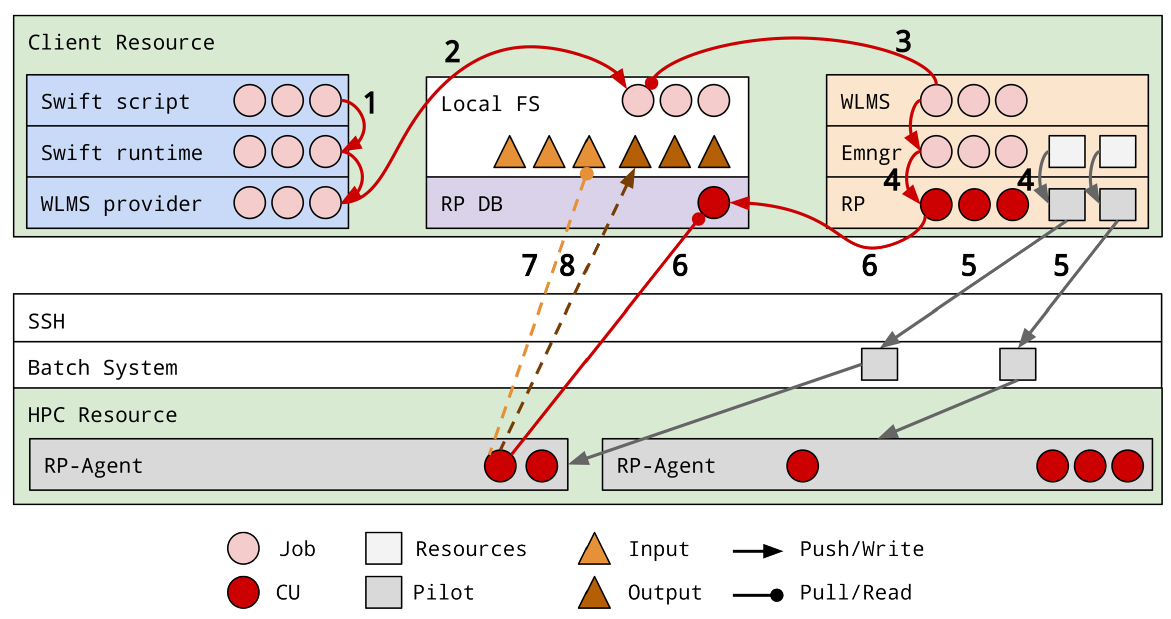}
    \caption{Integration between Swift and RADICAL-WLMS\@. The two systems
             exchange task descriptions via a local flilesystem. RADICAL-WLMS
             derives the size and duration of the pilots from the task
             requirements, independently from
             Swift.\label{fig:swift_integration}}
\end{figure}

We integrated Swift with RADICAL-WLMS to enable the distributed and
concurrent execution of Swift workflows on diverse resources
(Fig.~\ref{fig:swift_integration}). The execution strategies of RADICAL-WLMS
offered the possibility to minimize the time to completion of these
distributed executions, obtaining both qualitative and quantitative
improvements~\cite{turilli2016analysis}. Qualitatively, RADICAL-WLMS enabled
Swift to execute workflows concurrently on both HPC and HTC resources, via
late binding of both tasks to pilots and pilots to resources. Quantitatively,
the time to completion of workflows was improved by leveraging the shortest
queue time among all the target resources.

The integration with RADICAL-WLMS required the development of a dedicated
connector for Swift by iterating on the already available shell connector.
The RADICAL-WLMS connector enabled saving task descriptions on the local
filesystem from where RADICAL-WLMS was able to load and parse these
descriptions without needing any added functionality. This type of
integration was not made possible by an API---otherwise a common
implementation detail---but by sharing the task entity between the two
systems and by isolating distinct functionalities operating on that entity in
two distinct software modules.

Both Swift and RADICAL-WLMS are examples of building blocks for L3 (as
depicted in Fig.~\ref{fig:layers}) but most of their components are not. For
example, the workflow management component of Swift or the execution manager
component of RADICAL-WLMS are not designed to be self-sufficient and
extensible system that can be extended and composed with other building
blocks. Swift and RADICAL-WLMS' components can work only within those systems
because they depend on private APIs and assume specific coordination and
communication protocols.

\subsection{FireWorks and RADICAL-Pilot}\label{ssec:firework}

Fireworks is a workflow system with a large userbase and that enables
executing workflows on distributed and sometimes large scale compute
resources~\cite{jain2015fireworks}.

The design of FireWorks minimizes architectural and implementation complexity
while maximizing fault-tolerance and generality of workflow descriptions. The
system comprises four main components: a user-facing command-line tool to
describe workflows (lpad); a database where to store one or more workflows
(launchpad); a command-line tool to launch the execution of the workflows
(rlaunch); and a set of remote workers that execute the tasks of the
workflows on one or more resource (rockets).

When distributing the execution of workflows' tasks over resources, FireWorks
can benefit from late binding of tasks to resources. Nonetheless, FireWorks
does not implement pilot capabilities and therefore cannot late bind tasks on
HPC resources. This greatly reduces the potential of using HPC resources,
including the inability to support the high-throughput execution of MPI-based
simulations~\cite{angius2017converging}.

\begin{figure}
    \centering
    \includegraphics[width=0.49\textwidth]{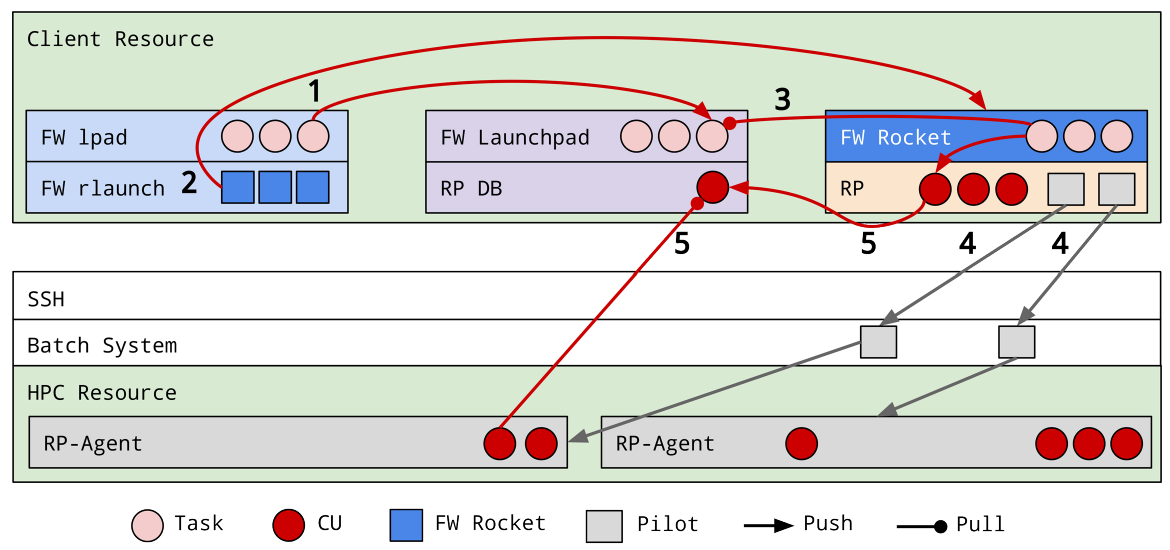}
    \caption{Integration between FireWorks (FW) and RADICAL-Pilot (RP). FW
             Rockets can be executed both locally or remotely. Rockets pass
             task descriptions and resource requirements to RP\@. RP works as
             a self-contained task execution
             system.\label{fig:fireworks_integration}}
\end{figure}

The integration of FireWorks with RADICAL-Pilot provides these pilot
capabilities. The two systems can be integrated at several levels (e.g., by
sharing their database or by replacing the existing FireWorks' workers) but
by enabling FireWorks' workers to submit jobs to RADICAL-Pilot
(Fig.~\ref{fig:fireworks_integration}), the isolation of states and the
assumptions behind FireWork's scheduling functionalities remain unaltered.

Unlike Swift, FireWorks does not offer a adapter subsystem but a worker can
be used to run a command via the RADICAL-Pilot API instead of a command to
immediately execute a task. In this way, RADICAL-Pilot behaves like an
independent subsystem that does not need to share any state but the initial
and final with FireWorks: Failures, rescheduling, resource selection, or the
multi-stage scheduling via pilots remain self-contained functionalities of
RADICAL-Pilot.

This case study confirms the `agnosticism' of modules designed as building
blocks towards API and integration points. Further, it also shows how single
modules of \rct can be integrated without the need to buy into the whole
stack. RADICAL-Pilot provides well-defined, self contained pilot capabilities
that can be integrated as-they-are into FireWorks, without requiring new
functionalities.

\subsection{PanDA and NGE}\label{ssec:panda}

PanDA is a Workload Management System designed to support the distributed
execution of workflows via pilots~\cite{maeno2008panda}. Pilot-capable WMS
enable high throughput of tasks execution via multi-level scheduling while
supporting interoperability across multiple sites. PanDA WMS consists of
several interconnected subsystems, communicating via dedicated API or HTTP
messaging and implemented by one or more modules.

PanDA is primarily designed to support execution of independent tasks on Grid
computing infrastructures like WLCG, but several prototypes have been
developed to support alternative platforms. Among these, leadership-class
computing systems are particularly promising as they typically run at 90\% of
their total yearly capacity. For example, the spare capacity of ORNL's Titan
supercomputer is equivalent to roughly 10\% of the 300,000 cores used by
PanDA on WLCG every year.

The use of leadership HPC machines for executing a very large amount of small
jobs presents several challenges. Among those, the two most relevant are:
coping with a queue system designed for large MPI jobs; and accessing the
untapped resources without disrupting the overall utilization of the machine.
Pilots can address the former while backfilling can be used for the latter.
Pilots can be difficult to deploy on HPCs because of the token-based
authentication model and the limited or absent WAN connectivity from the
compute nodes. Utilizing backfilling requires in turn dedicated development
and coordination with the management staff of the machine.

We developed and deployed a single-point solution to better understand the
problem space of enabling a workload management system designed for HTC to
execute production workflows on leadership-class resources designed to
support HPC\@. The PanDA team developed a job broker to support the execution
of part of the ATLAS production Monte Carlo workflow on Titan, while
RADICAL-Pilot and -SAGA were used to enable seamless pilot capabilities on
Titan via a new \rct we called Next Generation Executer (NGE).

\begin{figure}
    \centering
    \includegraphics[width=0.49\textwidth]{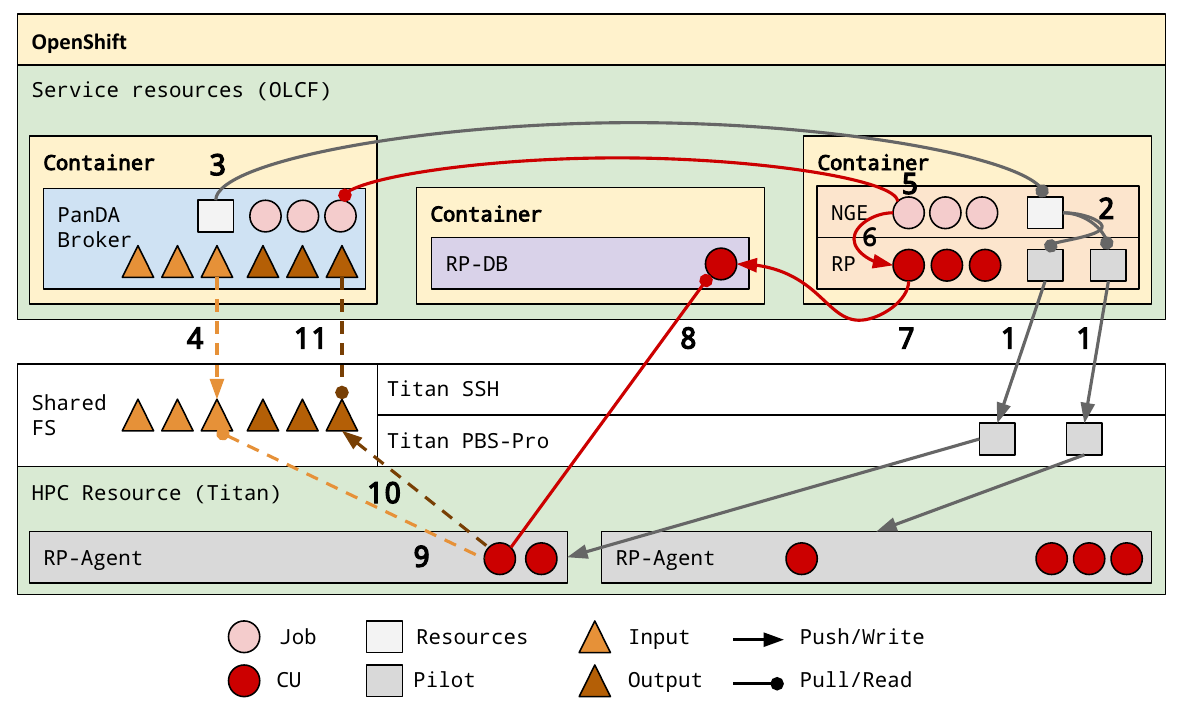}
    \caption{Integration between PanDA and Next Generation Executer (NGE).
             All systems execute on OLCF service resources within containers.
             Pilots are exposed to PanDA as an aggregation of available
             resources (steps 2 and 3).\label{fig:panda_integration}}
\end{figure}

PanDA Broker and NGE integrate via a database and a coordination protocol
based on exchanging information exclusively about tasks and resources
(Fig.~\ref{fig:panda_integration}). In this way, NGE behaves like a resource
queue for PanDA Broker while the broker is a source of tasks specifications
and resource requirements for NGE\@. Both systems require no modifications to
be integrated but the development of an API to pull and poll the database.

As with Swift and FireWorks, PanDA Broker is also developed by a dedicated
team, different from the one developing the \rct stack. The two teams did not
coordinate their design or development effort and the integration was
performed when the two stacks were already in production. Both systems
implement a design compatible with the building block approach, enabling
their integration. Nonetheless, their components are still tightly coupled
and interdependent: for example, the Agent of RADICAL-Pilot cannot be used in
isolation from the Unit and Pilot Managers and PanDA Broker cannot be used
without a PanDA Server and, to some extents, outside the boundaries of the
ATLAS experiment.

\subsection{Analysis}\label{ssec:analysis}

These four case studies show the potential for a set of software modules to
be designed without buying into the specific assumptions of a class of use
cases or types of resources. These assumptions have lead to several software
ecosystems that, while highly modular, do not allow reuse outside their
boundaries. We believe this is why functionalities pertaining to
domain-specific entities (e.g., tasks, pilots) are often reimplemented in use
case-specific software systems. Each system serves well the single research
group or the largest scientific project but not each other.

As argued in Section~\ref{sec:bba}, software modules should be
self-sufficient, interoperable, composabile, and extensibile so to be able to
serve a set of arbitrary requirements for a well-defined set of entities. For
example, a workflow manager should provide methods for DAG traversing
independent of how and when the DAG is specified or where the tasks of the
workflow will be executed. Analogously, a pilot agent module should provide
multi-staging and task execution capabilities independent on the system that
will schedule tasks on that agent or on the compute resources on which tasks
will be executed.

Modularity is not a design principle strong enough to realize this type of
software modules. Modularity needs to be augmented by API and coordination
agnosticism alongside an explicit understanding of the entities that define
the domain of utilization of the software system. Each module developed
following this approach, implements a well-defined set of functionalities
specific to a set of entities, with minimal assumptions about the system that
will use these functionalities or the environment in which they will be used.

This approach is by no means a design idealization or a complete novelty.
Systems like Celery, Dask, Kafka, or Docker are early examples of modules
designed by implicitly following what we have here called the building block
approach. These tools implement specific capabilities like queuing,
scheduling, streaming, or virtualization for the domain of distributed
computing. Consistently, they assume a set of core entities like concurrency,
workloads, tasks, pipelines, or messages. Their composability in multiple
domains and ongoing extensibility shows the potential of their underline
design approach.

\section{Discussion}\label{sec:discussion}

In Section~\ref{sec:bba} we described our interpretation of the building
blocks approach to design distributed systems and, in particular, workflow
systems. In Section~\ref{sec:rct} we illustrated how \rct were implemented in
accordance to this approach. Section~\ref{sec:casestudies} discussed the use
of \rct to constructing domain-specific workflow systems and integrating
legacy systems. There, we emphasized the ability to support a wide range of
scientific domains and applications with minimal perturbation and maximal
reuse of functionality and software.

This paper offers four main contributions: (i) Defining the principles of
self-sufficiency, interoperability, composability and extensibility that
characterize a building blocks approach to the design of distributed systems;
(ii) Illustrating a set of building blocks that enable multiple points of
integration, which results in design flexibility and functional
extensibility, as well as providing a level of ``unification'' in the
conceptual reasoning (e.g., execution) across otherwise very different tools
and systems; (iii) Showing how these building blocks have been used to
develop and integrate workflow systems; (iv) the beginning of an
investigation about an alternative and conceptual approach to (re)thinking
the design and implementation of workflow systems and the applicability and
potential of the building blocks approach.

The case studies we presented in Section~\ref{sec:casestudies} highlight the
practical impact of the building blocks approach. The first case study, based
on the integration of \entk and RADICAL-Pilot, illustrates how well-scoped
building blocks can support four domain-specific workflow systems, tailored
to the many distinct ensemble applications required by biomolecular sciences
and seismology. The three subsequent case studies show integration to provide
missing or improved functionality with widely used workflow systems and a
workload management system. The integration requires minimal development,
mainly focused on translation layers, and no refactoring. Together, these
four case studies meet the qualitative and quantitative requirements of a
variety of usage modes, a testament to the potential and impact of the
building block approach.

It is important to outline what this paper does not attempt to achieve. The
work captured in this work is not complete, in fact, it is a preliminary
study focused on one approach to building blocks for workflows systems,
without an encompassing analysis of application requirements. Although
preliminary, this work is not premature: Conceptual formalisms that are too
far ahead of proof-of-concepts and demonstrable advantages are unlikely to
yield practical advances. Thus, even though the building blocks approach is
still a work in progress, we believe early demonstrations of success are
necessary. Our work also does not attempt to distinguish (or identify) either
the set of applications or systems where a building blocks approach will
surpass alternative approaches. Finally, our paper does not analyze the wider
implications for the middleware ecosystem for scientific computing. We will
address these issues and more in future work.

The building blocks approach spawns many new questions. A prominent one
pertains to the issue of how we might model workflows systems and tools, so
as to provide a common vocabulary, reasoning and comparative framework. The
P* model provided this capability for the pilot
abstraction~\cite{review_pilotreview}, however it is still unclear what an
analogous conceptual model of workflow systems might entail or, given the
very broad diversity of workflow systems and tools, whether we can even
formulate a single conceptual model. This model has been elusive so far, but
might it be more fruitful to formulate a series of models of functional
modules that have the properties of building blocks as defined in Section 2?

There have been many surveys and analysis of workflow management
systems~\cite{atkinson-csur,deelman-fgcs,deelman_future_2017} which have
focused on a functional analysis of workflows and classification of workflows
systems. To the best of our understanding, a survey that has examined
workflow systems from a software engineering perspective and practice is
conspicuous by its absence. An end-goal and intended outcome of this paper is
to begin a discussion on how the scientific workflows community---end-users,
workflow designers when distinct from end-users, and workflow systems
developers---can better coordinate, cooperate, and reduce redundant and
unsustainable efforts. We believe the building blocks approach is a start
towards an examination and investigation of design principles and
architectural patterns for workflow systems that may facilitate this
discussion.

\section{Acknowledgments}

This work is made possible by the work of Vivek Balasubrmanian (\entk) and
other RADICAL members. We thank our collaborators: Peter Coveney, Dave Wright
and Stefan Zasada (UCL/HTBAC); Ruslan Mashnitov, Danila Oleynik, Kaushik De,
Jack Well and Alexei Klimentov (ATLAS Project/PanDA); Matthieu Lefevbre, Ryan
Modrak, Wenjie Lie and Jeroen Tromp (Princeton/SeisFlows); Charlie Laughton
(Nottingham) and Cecilia Clementi (Rice/ExTASY); Anubhav Jain and Joseph
Montoya(Berkeley Labs/Fireworks). We thank NSF and DOE for financial support,
and  PRAC (Blue Waters) and XRAC awards for Computational Resources.

\bibliographystyle{abbrv}
\bibliography{works17}

\end{document}